\documentclass[11pt]{article}
\usepackage[utf8]{inputenc}

\usepackage{jheppub}

\usepackage[capitalize]{cleveref}
\usepackage{physics}

\newcommand{\as}{{\alpha_S}}
\newcommand{\aso}{{\alpha_{S,0}}}

\title{NNLO QCD corrections to event shapes at the LHC}

\author[c]{Manuel Alvarez,}
\author[d]{Josu Cantero,}
\author[a]{Michal Czakon,}
\author[e]{Javier Llorente,}
\author[b]{Alexander Mitov,}
\author[b]{Rene Poncelet}

\affiliation[a]{Institut f\"ur Theoretische Teilchenphysik und Kosmologie, RWTH Aachen University,\\ D-52056 Aachen, Germany}
\affiliation[b]{Cavendish Laboratory, University of Cambridge, Cambridge CB3 0HE, United Kingdom}
\affiliation[c]{Departamento de Física Teórica, Universidad Autónoma de Madrid. Madrid, Spain.}
\affiliation[d]{Instituto de Física Corpuscular. Valencia, Spain.}
\affiliation[e]{Department of Physics, Simon Fraser University. Burnaby, Canada.}

\emailAdd{mczakon@physik.rwth-aachen.de}
\emailAdd{adm74@cam.ac.uk}
\emailAdd{poncelet@hep.phy.cam.ac.uk}
\emailAdd{josu.cantero.garcia@cern.ch}
\emailAdd{manuel.alvarez.estevez@cern.ch}
\emailAdd{javier.llorente.merino@cern.ch}

\preprint{Cavendish-HEP-22/11, P3H-22-129, TTK-22-49}

\abstract{In this work we perform the first ever calculation of jet event shapes at hadron colliders at next-to-next-to leading order (NNLO) in QCD. The inclusion of higher order corrections removes the shape difference observed between data and next-to-leading order predictions. The theory uncertainty at NNLO is comparable to, or slightly larger than, existing measurements. Except for narrow kinematical ranges where all-order resummation becomes important, the NNLO predictions for the event shapes considered in the present work are reliable. As a prime application of the results derived in this work we provide a detailed investigation of the prospects for the precision determination of the strong coupling constant and its running through TeV scales from LHC data.}

\begin{document}
\maketitle
\flushbottom

\section{Introduction}
\label{sec:introduction}

One of the most direct ways of exhibiting the QCD dynamics at hadron colliders is by studying multi-jet events. Such events are special since they are not very sensitive to the electroweak (EW) sector of the Standard Model (SM), contain almost exclusively QCD partons, and are driven by QCD's dynamics, color structure and single coupling constant, $\as$.

The characterization of multi-jet events is often performed in terms of so-called event shapes. These are single-valued observables which encode the events' topology and energy flow. Event shapes are constructed in such a way that configurations with back-to-back jets can naturally be singled out from ones with more isotropic energy distribution in the final state. Event shapes are sensitive to the probability of additional emissions, i.e., the QCD dynamics and the strong coupling constant. 

First measurements of multi-jet event shapes were performed at LEP \cite{ALEPH:2003obs, OPAL:2004wof, DELPHI:2003yqh, L3:2004cdh} and were used to extract $\as$ \cite{Kluth:2000km}. Event shapes and multi-jet production at hadron colliders have been studied extensively, starting with the Tevatron \cite{CDF:2011yfm}. At the LHC, the ATLAS and CMS experiments have measured generic multi-jet differential cross-sections \cite{ATLAS:2011qvj, CMS:2013vbb, CMS:2014mna, ATLAS:2014qmg}, event shapes at 7 and 13 TeV~\cite{ATLAS:2012tch, CMS:2013lua, CMS:2014tkl, ATLAS:2015yaa,CMS:2018svp, ATLAS:2020vup}, azimuthal decorrelations at 8 TeV \cite{ATLAS:2018sjf} and transverse-energy-energy-correlations at 13 TeV \cite{ATLAS:2020mee}.

Theoretical work on jet production in hadron-hadron collisions also has a long history. Fixed-order, next-to-leading-order (NLO) QCD corrections for dijet events have been first obtained in refs.~\cite{Ellis:1992en, Giele:1993dj}. These have been complemented by NLO EW corrections \cite{Dittmaier:2012kx, Frederix:2016ost} and parton-shower effects \cite{Alioli:2010xa, Hoeche:2012fm}. Dijet cross-sections with NLO accuracy are also available from general-purpose event generators \cite{Gleisberg:2008ta, Alwall:2014hca, Frederix:2018nkq}. The computation of next-to-next-to-leading-order (NNLO) QCD corrections to dijet events has been achieved relatively recently~\cite{Currie:2016bfm, Currie:2017eqf, Czakon:2019tmo}. 

Based on simple energy-momentum considerations it is clear that dijet events contribute trivially to event shapes. Therefore, only events with three or more jets in the final state can span non-trivial values of event shapes while dijet events tend to only have endpoint contributions. Three-jet cross-sections have been studied at NLO QCD \cite{Nagy:2001fj, Nagy:2003tz} and NLO EW, including parton-shower effects \cite{Reyer:2019obz}. Very recently, the first computation of NNLO QCD corrections to three-jet production has been performed by some of the authors of the present article \cite{Czakon:2021mjy}, mainly focusing on three-to-two jet ratios. Higher jet multiplicities have also been investigated in the literature with up to five jet production known through NLO in QCD~\cite{Bern:2011ep, Badger:2013yda}.

Fixed-order calculations do not always adequately describe multi-jet event shapes. For certain kinematic configurations one has to resum contributions from QCD emissions to all orders in $\as$ in order to have sensible theoretical predictions. Resummation with Next-to-leading-logarithmic (NLL) accuracy matched to NLO QCD predictions has been achieved for global event shapes \cite{Banfi:2004nk, Banfi:2010xy}. As their name suggests, global event shapes feature the important property that they are defined in the full phase space. The presence of phase space restriction like phase space cuts, radiation gaps, etc., renders an event shape non-global and gives rise to associated logarithms, so-called non-global logarithms \cite{Dasgupta:2001sh, Banfi:2010pa}. The resummation of non-global logarithms is notoriously difficult and has been achieved only in special cases, see for example, refs.~\cite{Khelifa-Kerfa:2011quw, Tackmann:2012bt, Larkoski:2015zka, AngelesMartinez:2018cfz, Michel:2018hui, Banfi:2021owj, Becher:2021urs}. 

The full phase space is never accessible at hadron colliders due to the missing forward coverage of the detector. This limits the implementation of global event shapes experimentally. A possible work-around is to re-define event shapes by exponentially suppressing recoil terms, which regulates the contribution from the unavailable phase space \cite{Banfi:2010xy}. However, such an implementation would require the measurement of event's hadronic recoil which is accompanied by a significant experimental uncertainty. It also requires the measurement of individual particle momenta which increases the sensitivity to hadronisation and other non-perturbative effects that are difficult to control.

An alternative approach to event shapes is to define event shapes in terms of reconstructed jets instead of particles. Such an approach has the advantage that non-perturbative effects from hadronisation and multi-parton interactions (MPI) are suppressed. Experimentally, working with jets instead of calorimeter clusters, which are the proxies of the particles at the detector level, is also advantageous since detector-level jets are well-defined and well-calibrated objects. The calibration of the jet energy scale and resolution together with their corresponding uncertainties is typically well under control, whereas this not the case for calorimeter clusters in general. In addition, it is easier to associate detector-level jets to particle-level jets, in contrast to the case of a calorimeter cluster which typically cannot be uniquely associated to a single particle.

Event shapes are a powerful tool for testing parton-shower models in state-of-the-art Monte Carlo simulations and for tuning shower parameters. For example, differences in the matching to fixed-order matrix elements and in the ordering of the emissions ($p_T, \theta$,  etc) can lead to important differences in the description of these observables~\cite{ATLAS:2020vup}. 

In this work, we compute the NNLO QCD corrections to a set of event shapes based on reconstructed jets. As explained above, jet-based event shapes allow for precise experimental measurements, i.e. an increased precision of theoretical predictions will make it possible to closely scrutinize our ability to precisely describe QCD at hadron colliders and to extract $\as$ and its running through TeV scales with unprecedented precision. While in this study we do not use MC or resummed predictions, we identify regions where such corrections might be needed.

This article is organised as follows: in sec.~\ref{sec:comp} we give details about the computational setup and the definitions of the event shapes studied in the present work. The main results are presented in sec.~\ref{sec:eventshapes}. Specifically, perturbative corrections and PDF effects are discussed in sec.~\ref{sec:eventshapes-pQCD}, the dependence on $\as$ and its extraction in sec.~\ref{sec:eventshapes-alphas}, while sec.~\ref{sec:eventshapes-np} contains an estimation of non-perturbative effects. Our conclusions can be found in sec.~\ref{sec:conclusion}.

\section{Definitions and computational details}\label{sec:comp}

In this work we compute the NNLO QCD corrections to the set of multi-jet event shape observables specified in sec.~\ref{sec:comp-eventshapes} below. To span the full kinematic ranges of these observables, final states containing at least three jets are required. For the calculation of the NNLO QCD corrections to the inclusive three-jet cross section we employ the sector-improved residue subtraction framework \cite{Czakon:2010td, Czakon:2014oma}, which has already been applied to inclusive jet~\cite{Czakon:2019tmo}, two jet- as well three-jet production \cite{Czakon:2021mjy}. Tree-level matrix elements are evaluated with the help of the {\tt AvH} library~\cite{Bury:2015dla}. All contributing one-loop amplitudes are taken from the {\tt OpenLoops 2} library \cite{Buccioni:2019sur}. The two-loop amplitudes for three-jet production are only available in the leading-colour approximation~\cite{Abreu:2019odu}. In the present work we utilize these approximate double virtual corrections to three-jet production along the lines of ref.~\cite{Czakon:2021mjy}. Specifically, the two-loop finite remainder
\begin{eqnarray}
 \mathcal{R}^{(2)} (\mu_R^2) &=&
   2 \Re\left[ \mathcal{M}^{\dagger (0)}\mathcal{F}^{(2)}\right] (\mu_R^2)
   + \big\vert \mathcal{F}^{(1)}\big\vert^2(\mu_R^2)\nonumber\\
   &\equiv & {\cal R}^{(2)}(s_{12})+\sum_{i=1}^4 c_i
                           \ln^i\left(\frac{\mu_R^2}{s_{12}}\right)\,,
\end{eqnarray}
is approximated by replacing the function ${\cal R}^{(2)}(s_{12})$ with its leading-colour approximation provided with the software described in ref.~\cite{Abreu:2021oya}. It is implemented in terms of the so-called ``pentagon functions" \cite{Chicherin:2020oor} and rational functions of the kinematic invariants \cite{Abreu:2019odu}. Similar to the results presented in ref.~\cite{Czakon:2021mjy}, sub-leading colour corrections are expected to be of order $1\%$ with respect to the differential three-jet cross sections. We work in the $n_f=5$ scheme, i.e. we consistently neglect contributions from top-quarks in the partonic cross sections and in the running of $\as$.

The $\as$ expansion of the $n$-jet differential cross section reads:
\begin{align}
\dd \sigma_n(\mu_R,\mu_F,\text{PDF},\as(\mu_R,\aso)) &= \sum_i \as^{n+i}(\mu_R,\aso) \dd \sigma^{(i)}_n(\mu_R,\mu_F,\text{PDF})\,,
\end{align}
where $\mu_{R,F}$ are the renormalisation and factorisation scales and PDF labels the parton distribution set. The dependence on the initial value of the strong coupling $\aso \equiv \as(\mu_R = m_Z)$ is also made explicit for later convenience.

The strong coupling constant satisfies the following renormalisation group equation (RGE):
\begin{align}
    \mu_R^2\frac{\dd \as}{\dd \mu_R^2} = \beta(\as) = -(b_0 \as^2 + b_1 \as^3 + b_2 \as^4 + \dots)\,,
    \label{eq:as_rge}
\end{align}
where $b_{0,1,2}$ are the 1-, 2- and 3-loop $\beta$-function coefficients. The strong coupling constant at a scale $\mu_R$, together with the corresponding PDF set, is obtained from the {\tt LHAPDF} library \cite{Buckley:2014ana}. Specifically, $\as(\mu_R)$ is derived from the numerical solution of the RGE eq.~(\ref{eq:as_rge}) starting from the reference scale $\mu_R=m_Z$. In the present work we are interested in the three lowest orders in perturbation theory and define them for convenience as
\begin{align}
    \dd \sigma_n^{\text{LO}}   &= \as^n \dd \sigma^{(0)}_n\,, \nonumber\\
    \dd \sigma_n^{\text{NLO}}  &= \as^n \dd \sigma^{(0)}_n + \as^{n+1} \dd \sigma^{(1)}_n\,,\nonumber\\
    \dd \sigma_n^{\text{NNLO}} &= \as^n \dd \sigma^{(0)}_n + \as^{n+1} \dd \sigma^{(1)}_n + \as^{n+2} \dd \sigma^{(2)}_n\,.
\end{align}

All multi-jet observables in this work are computed as ratios
\begin{align}
    R^i(\mu_R,\mu_F,\text{PDF},\aso) = \frac{\dd \sigma^i_3(\mu_R,\mu_F,\text{PDF},\aso)}
                                             {\dd \sigma^i_2(\mu_R,\mu_F,\text{PDF},\aso)}\,,
\label{eq:ratio_def}
\end{align}
for $i \in \{\text{LO, NLO, NNLO}\}$. We study the NLO and NNLO QCD $\mathcal{K}$-factors:
\begin{align}
    \mathcal{K}^{\text{NLO}} = \frac{R^{\text{NLO}}}{R^{\text{LO}}} \quad \text{and} \quad \mathcal{K}^{\text{NNLO}} = \frac{R^{\text{NNLO}}}{R^{\text{NLO}}}\;.
\end{align}
To estimate missing higher-order contributions, we take the envelope of a 7-point variation of $\mu_R$ and $\mu_F$ by a factor of $2$ with the constraint $\frac{1}{2} \leq \frac{\mu_R}{\mu_F} \leq 2$. For all predictions presented in this work, we use as a central scale choice
\begin{align}
    \mu_R = \mu_F = \hat{H}_T = \sum_{i\in \text{partons}} p_{T,i}\,,
    \label{eq:def_scale}
\end{align}
where the sum runs over all partons in the final state.

The PDF uncertainties are estimated by following the prescriptions of the individual PDF sets. Due to the high computational costs, PDF member variations have been evaluated only at NLO QCD using the NNLO version of the corresponding PDF set, and then extrapolated to NNLO QCD by using the following $\mathcal{K}$-factor approach
\begin{align}
R^{\text{NNLO}}(\text{PDF}_i) \approx \frac{R^{\text{NLO}}(\text{PDF}_i)}{R^{\text{NLO}}(\text{PDF}_0)} \;R^{\text{NNLO}}(\text{PDF}_0)\,,
\end{align}
where $\text{PDF}_i$ denotes the $i$-th member of the PDF set ($0$ corresponds to the central member). The justification for this approximation is that the ratios have a reduced dependence on the PDF due to cancellation between the two- and three-jet cross sections. Indeed the PDF uncertainties for the ratios are typically small and dwarfed by the remaining numerical integration uncertainties.

\subsection{Definition of event shapes}\label{sec:comp-eventshapes}

This section provides the definitions of all event shapes computed in section \ref{sec:eventshapes}.

A classical event shape observable is the transverse thrust $T_{\perp}$ (or rather $\tau_{\perp} = 1- T_{\perp}$) \cite{Brandt:1964sa, Farhi:1977sg} and its minor component $T_m$, defined by
\begin{align}
    T_\perp = \frac{\sum_i | \vec{p}_{T,i} \cdot \hat{n}_{\perp}|}{\sum_i | \vec{p}_{T,i}|}\;, \quad \text{and} \quad T_m = \frac{\sum_i | \vec{p}_{T,i} \times \hat{n}_{\perp}|}{\sum_i | \vec{p}_{T,i}|}\;.
\label{eq:T-def}    
\end{align}

The thrust axis $n_{\perp}$ maximises the projection of all jets to this axis. As explained in the Introduction, we define the thrust and all other event shapes in terms of reconstructed jets, i.e. the sum $i$ in eq.~(\ref{eq:T-def}) above runs over the list of all reconstructed jets passing selection criteria. The thrust separates topologies that are of back-to-back type (small $\tau_{\perp}$) from ones that are isotropic (large $\tau_{\perp})$. Another event shape that characterises the anisotropy of an event is the linearised sphericity tensor \cite{Parisi:1978eg, Donoghue:1979vi}:
\begin{align}
    \mathcal{M}_{xyz} = \frac{1}{\sum_i |\vec{p}_i|} \sum_i \frac{1}{|\vec{p}_i|}
    \left(
      \begin{array}{ccc}
       p_{x,i}^2      & p_{x,i}p_{y,i} & p_{x,i} p_{z,i} \\
       p_{y,i}p_{x,i} & p_{y,i}^2      & p_{y,i} p_{z,i} \\
       p_{z,i}p_{x,i} & p_{z,i}p_{y,i} & p^2_{z,i}  %
    \end{array}
    \right)\,.
\end{align}
In this work, we study certain combinations of the eigenvalues $\lambda_1,\lambda_2,\lambda_3$ of $\mathcal{M}_{xyz}$. They are denoted $A$ (the so-called aplanarity), $C$, and $D$ \cite{Ellis:1980wv}:
\begin{align}
    A = \frac{3}{2}\lambda_3\;,\quad
    C = 3 (\lambda_1 \lambda_2 + \lambda_1 \lambda_3 + \lambda_2\lambda_3)\;,\quad
    D = 27 \lambda_1 \lambda_2 \lambda_3\;.
\end{align}
Furthermore, from the transverse linearised sphericity tensor 
\begin{align}
    \mathcal{M}_{xy} = \frac{1}{\sum_i |\vec{p}_{T,i}|} \sum_i \frac{1}{|\vec{p}_{T,i}|}
    \left(
      \begin{array}{cc}
       p_{x,i}^2      & p_{x,i}p_{y,i} \\
       p_{y,i}p_{x,i} & p_{y,i}^2
    \end{array}
    \right)\;,
\end{align}
with eigenvalues $\mu_1,\mu_2$, one defines the transverse sphericity variable:
\begin{align}
    S_\perp = \frac{\mu_2}{\mu_1+\mu_2}\;.
\end{align}

The final observable we study in this article is the transverse energy-energy correlator (TEEC) \cite{CERN-TH-3800, TEEC-TH-NLO, doi:10.1142/2300, ATLAS:2017qir, ATLAS:2020mee}. For multi-jet events, it is defined as
\begin{align}
  \frac{1}{\sigma_2}\frac{\dd \sigma}{\dd \cos\Delta\phi}
   = \frac{1}{\sigma_2}
   \sum_{ij}\int \frac{\dd \sigma\; x_{\perp,i} x_{\perp,j}}{\dd x_{\perp,i} \dd x_{\perp,j} \dd \cos\Delta\phi_{ij}}
    \delta(\cos\Delta\phi-\cos\Delta\phi_{ij}) \dd x_{\perp,i} \dd x_{\perp,j} \dd \cos\Delta\phi_{ij} \,,
\label{eq:TEEC}    
\end{align}
where $x_{\perp, i} = E_{\perp, i}/\sum_k E_{\perp,k}$, and with $E_{\perp} = \sqrt{E^2-p_z^2}$ being invariant under boosts along the z-axis. The angle $\Delta\phi_{ij}$ is the azimuthal opening between a pair of jets $ij$. 

Note that in the presence of phase space cut(s), the TEEC picks up an implicit dependence on the corresponding kinematic variable(s). As specified in sec.~\ref{sec:eventshapes} below, the TEEC is measured in bins of the variable $H_{T,2} $. The denominator in eq.~(\ref{eq:TEEC}) is the two-jet cross-section evaluated in the corresponding bin of $H_{T,2} $. To be able to generate a contribution away from the endpoints $\cos\Delta\phi_{ij} = \pm 1$, the numerator in eq.~(\ref{eq:TEEC}) requires at least a $2 \to 3$ process. A fixed-order $2 \to 3 $ calculation diverges at the endpoints $\cos\Delta\phi_{ij} = \pm 1$ which correspond to the back-to-back limit. To avoid these infrared sensitive regions, a constraint $|\cos\Delta\phi_{ij}|\leq |\cos\Delta\phi_{\text{max}}|$ is imposed. For a sufficiently small value of $|\cos\Delta\phi_{\text{max}}|$ contributions from resummation-dominated phase space regions are suppressed and fixed-order perturbative calculations are expected to be reliable.

\section{Event-shapes at the LHC}\label{sec:eventshapes}

In this section we present NNLO QCD predictions for event shape observables measured at 13 TeV by the ATLAS collaboration \cite{ATLAS:2020vup}
\footnote{The ATLAS measurement has been performed in jet-multiplicity bins. We compare our predictions with the inclusive data presented in appendix A.}. 
For the numerator (denominator) of eq.~(\ref{eq:ratio_def}) we require at least three (two) $R=0.4$ anti-$k_T$ \cite{Cacciari:2008gp} jets that fulfill the requirements
\begin{itemize}
\item $H_{T,2} = p_{T,1}+p_{T,2} \geq 1~\text{TeV}$ where $p_{T,i}$ is the transverse momentum of the $i$-th hardest jet,
\item $|y(j)| \leq 2.4$ and $p_T(j) \geq 100~\text{GeV}$.
\end{itemize}
The inputs for the event shapes are all jets that pass the above requirements.

Due to the high computational cost of the three jet cross section, we compute the event shapes in bins that are coarser than the bins used for the measurement in ref.~\cite{ATLAS:2020vup}. To facilitate the comparison between the measurement and our predictions we rebin the data as follows: the systematic uncertainties between bins are considered as fully correlated, whereas the statistical uncertainties are added in quadrature. All event shapes are measured in three separate $H_{T,2}$ regions
\begin{align*}
    1000~\text{GeV} \leq H_{T,2} < 1500~\text{GeV},\quad
    1500~\text{GeV} \leq H_{T,2} < 2000~\text{GeV},\quad
    2000~\text{GeV} \leq H_{T,2}\;,
\end{align*}
which allows for the separation of different energy scales.

For the computation of the TEEC, see eq.~(\ref{eq:TEEC}), we adopt a slightly different phase space corresponding to the selection in ref.~\cite{ATLAS:2020mee}:
\begin{itemize}
\item $H_{T,2} = p_{T,1} + p_{T,2} \geq 1~\text{TeV}$,
\item at least 2 anti-$k_T$ jets with $R=0.4$ and $p_T \geq 60~\text{GeV}$ $|y(j)| \leq 2.4$.
\end{itemize}
For this observable we consider a finer $H_{T,2}$ binning
\begin{align*}
    [1000,1200,1400,1600,1800,2000,2300,2600,3000,3500,\infty)\;,
\end{align*}
together with the ``inclusive" case $H_{T,2} \geq 1000~\text{GeV}$.

The central choice for renormalisation and factorisation scales is $\hat{H}_T$, defined in equation (\ref{eq:def_scale}). As the default PDF choice, we use the \textsc{NNPDF30}\cite{NNPDF:2014otw} PDF set
\footnote{This PDF set has been superseded by \textsc{NNPDF31}~\cite{NNPDF:2017mvq} and \textsc{NNPDF40}~\cite{NNPDF:2021njg} which have smaller uncertainties and are based on better methodology. These differences are not so crucial for the current discussion as the PDF uncertainties are comparatively small.}.

\subsection{Perturbative QCD corrections}\label{sec:eventshapes-pQCD}

\begin{figure}
    \centering
    \includegraphics[page=1,width=0.5\textwidth]{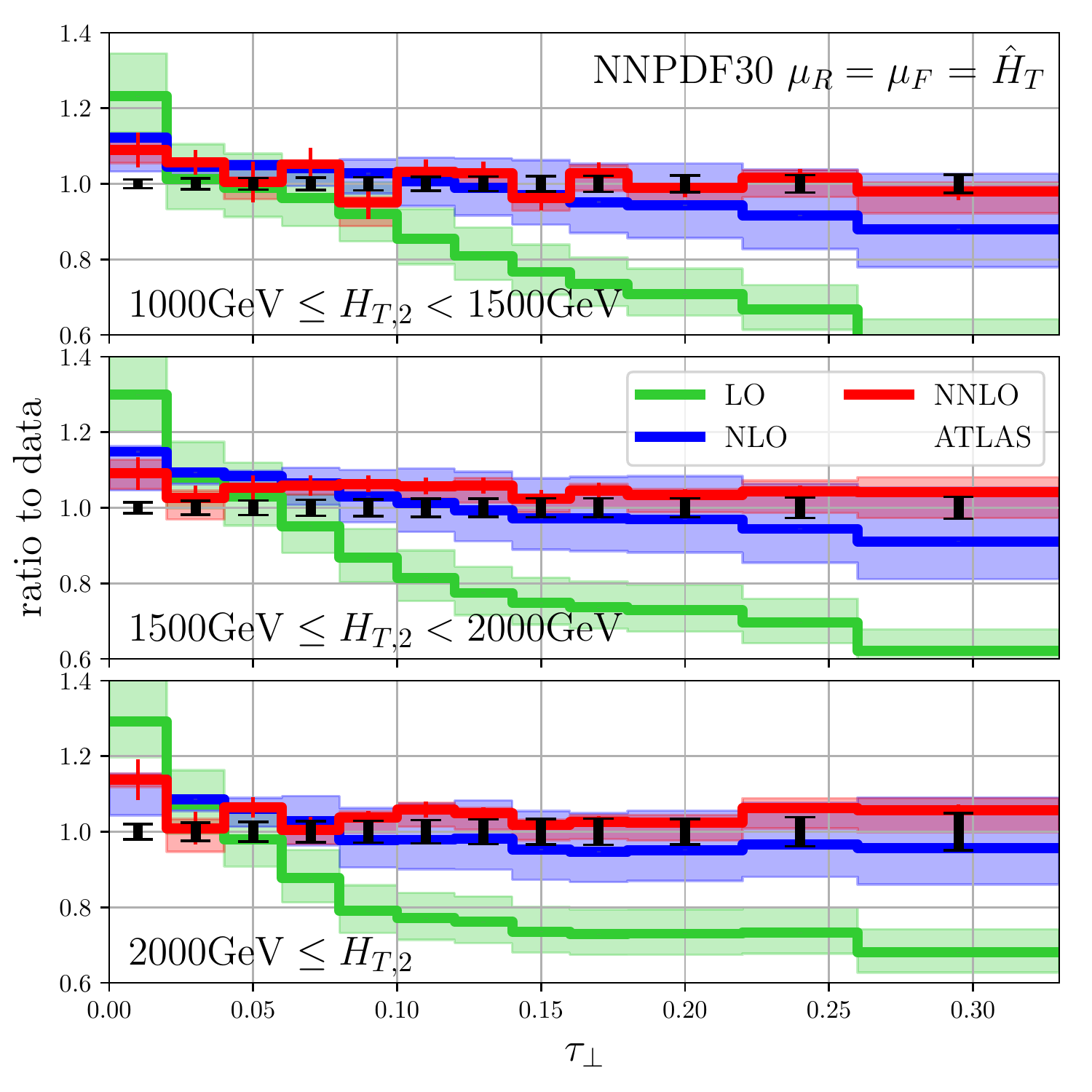}%
    \includegraphics[page=2,width=0.5\textwidth]{plots/NNPDF30-HThat-r32-data.pdf}
    \caption{
    The transverse thrust $\tau_\perp$ (left) and the thrust minor $T_m$ (right) in three $H_{T,2}$ bins.
    The solid lines show fixed order LO (green), NLO (blue), and NNLO (red) predictions normalised to ATLAS data (black) \cite{ATLAS:2020vup}.
    The coloured bands show the scale variation.}
    \label{fig:eventshapes-pQCD-1}
\end{figure}

\begin{figure}
    \centering
    \includegraphics[page=5,width=0.5\textwidth]{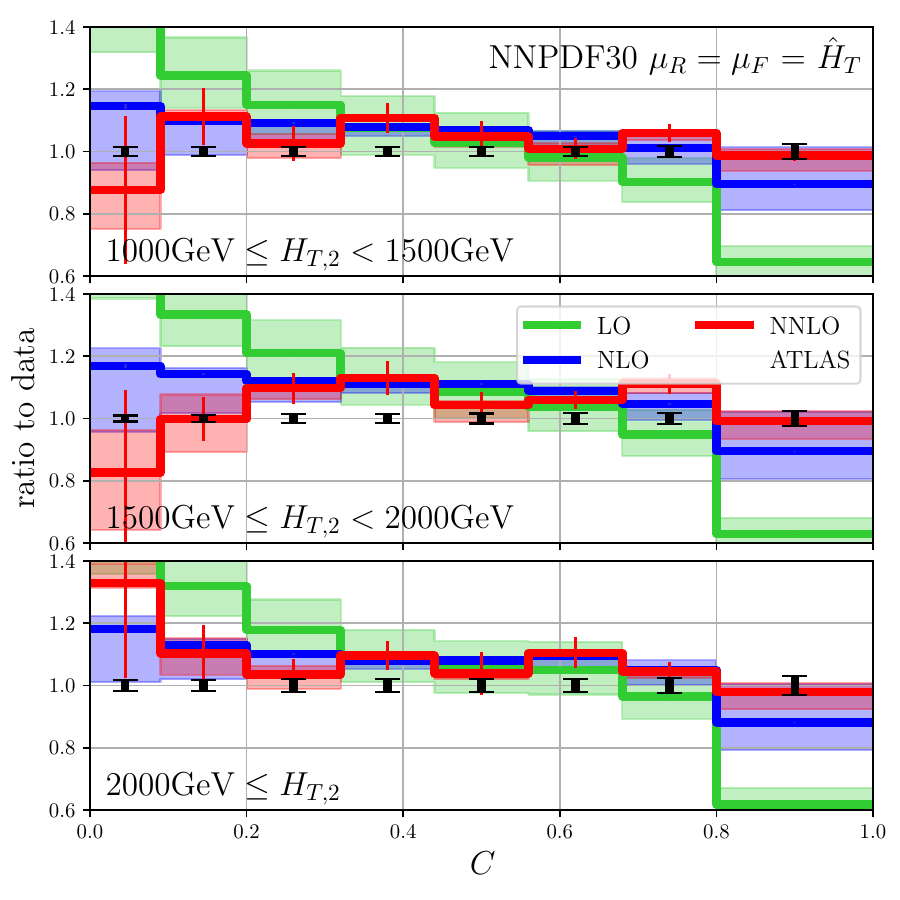}%
    \includegraphics[page=6,width=0.5\textwidth]{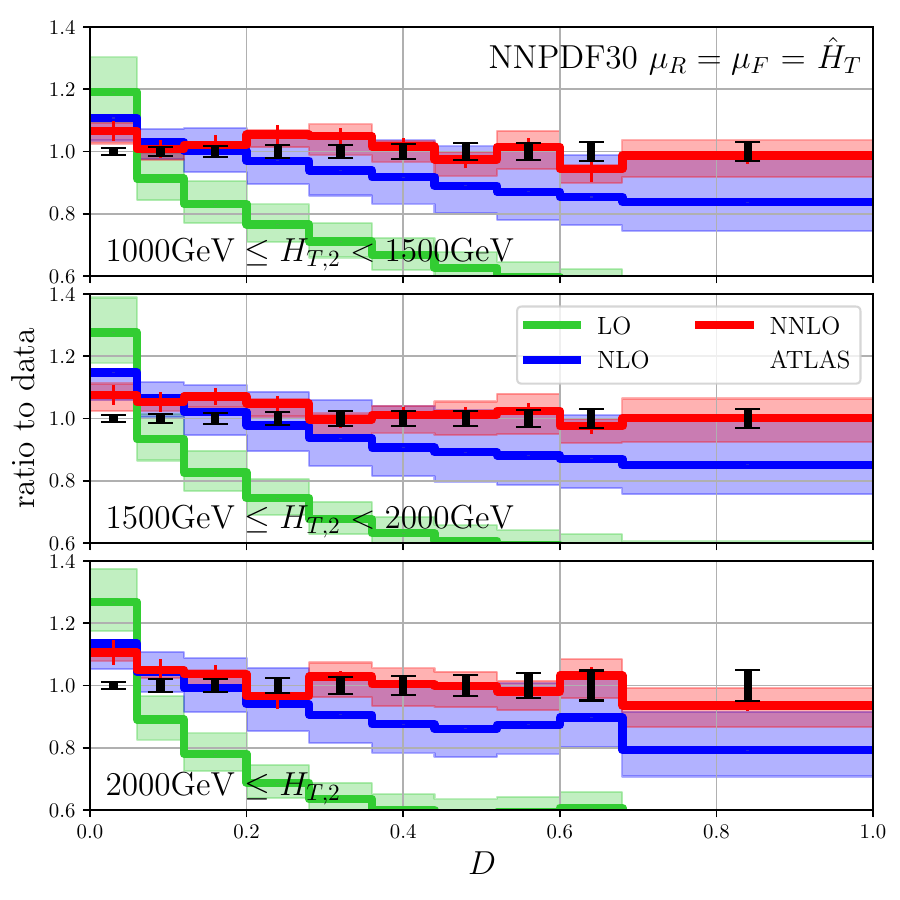}
    \caption{
    As in fig.~\ref{fig:eventshapes-pQCD-1} but for the variables $C$ (left) and $D$ (right).}
    \label{fig:eventshapes-pQCD-3}
\end{figure}

In fig.~\ref{fig:eventshapes-pQCD-1} to fig.~\ref{fig:eventshapes-pQCD-3}, we show the LO, NLO and NNLO QCD corrections to the six event shapes defined in section \ref{sec:comp-eventshapes}. All predictions are shown as ratios to the data. Coloured bands show the envelope of the 7-point scale variation, and vertical bars indicate the remaining uncertainty from Monte Carlo integration. The data is shown with a bar indicating the statistical and systematical uncertainty added in quadrature.

The results for the transverse thrust $\tau_\perp$ are shown on the left-hand side of fig.~\ref{fig:eventshapes-pQCD-1}. We observe significant NLO QCD corrections to the shape of the LO prediction. In particular, for large thrust values representing events with hard resolved radiation, corrections can be as large as $50\%$. The scale uncertainty increases, indicating that the LO scale dependence does not suffice to estimate missing higher effects for this observable. The NLO result does agree fairly well with data within the uncertainties, but one can observe the tendency to undershoot the data in the tail of the distribution. Turning to the NNLO QCD results, we observe small corrections to the shape at NLO and a sizable reduction of the scale dependence by a factor of two to four. The NNLO QCD predictions are fully contained within the NLO QCD scale uncertainties indicating perturbative convergence. Agreement with data, within uncertainties, can be observed over the full range of the thrust observable. Importantly, the NNLO QCD corrections lead to theory/data shape agreement in the tail. 

The behavior of the transverse thrust $\tau_\perp$ for small values of $\tau_\perp$ deserves special attention. As can be observed in fig.~\ref{fig:eventshapes-pQCD-1} the behavior of the fixed order prediction at all three orders is different, showing change in slope and increased scale variation, especially at LO. This behavior is accompanied by an increase in the MC uncertainty of the predictions and the agreement with data at NNLO is worse than in all other bins. This is the case for all three $H_{T,2}$ slices. The reason behind this behavior is that in this kinematics the fixed order expansion starts to break down and all-order resummation effects become important. This behavior is well known from the $e^+e^-$ thrust \cite{Becher:2008cf} for which very high resummation as well as fixed order accuracy has been achieved. In general, we wish to point out the substantial progress that has recently been achieved for the resummation and the description of event shape observables at high energy colliders \cite{Tulipant:2017ybb, Moult:2018jjd, Gao:2019ojf, Moult:2019mog, Gehrmann:2019hwf, Bris:2020uyb, Ebert:2020sfi, Beneke:2022obx}.

On the right-hand side of fig.~\ref{fig:eventshapes-pQCD-1}, we show the thrust minor component $T_m$. The pattern of higher-order corrections is similar to the thrust observable: large NLO and smaller NNLO corrections. The reduction in scale dependence when going from NLO to NNLO QCD, particularly for larger values of $T_m$, is large and can reach a factor of 10 in the intermediate region. The NLO QCD predictions for $T_m$ do not describe the shape of the observable very well, especially around the peak region $T_m\sim0.15$. This situation is improved by the NNLO QCD calculation, which predicts the shape well. The lowest $T_m$ bin suffers from large statistical uncertainties on the NNLO QCD predictions, for reasons similar to the ones discussed in the context of the transverse thrust $\tau_\perp$.

\begin{figure}
    \centering
    \includegraphics[page=3,width=0.5\textwidth]{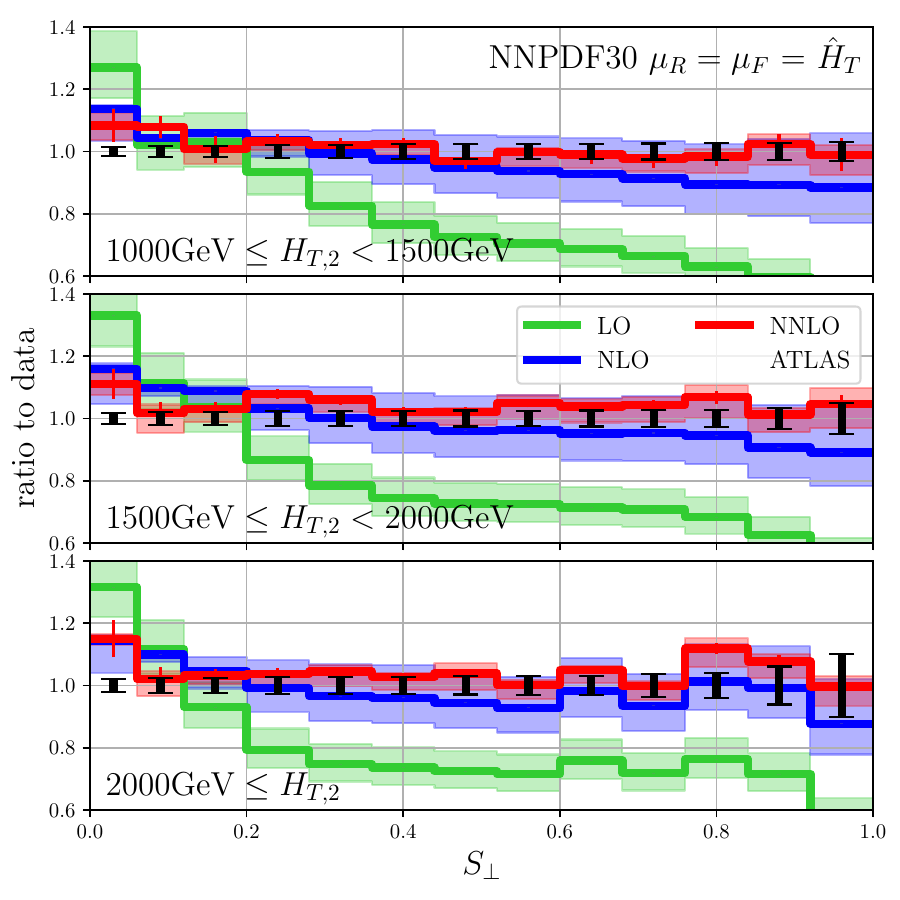}%
    \includegraphics[page=4,width=0.5\textwidth]{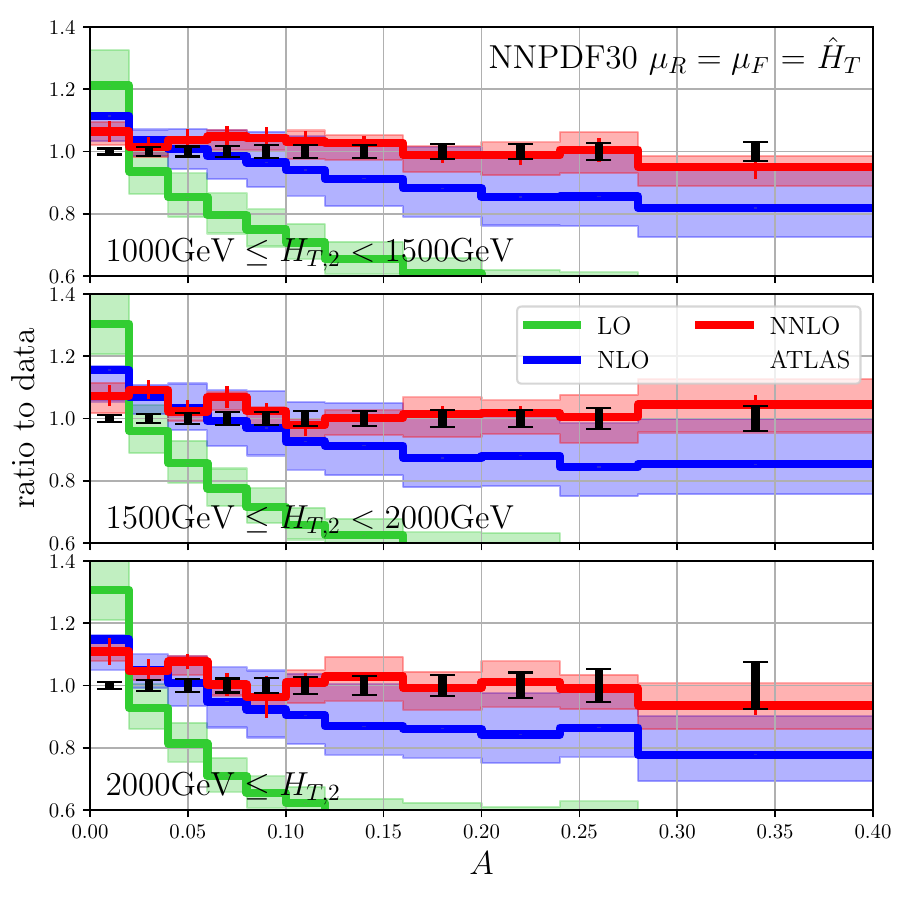}
    \caption{
    As in fig.~\ref{fig:eventshapes-pQCD-1} but for the transverse sphericity $S_{\perp}$ (left) and the aplanarity $A$ (right).}
    \label{fig:eventshapes-pQCD-2}
\end{figure}

We find good agreement between ATLAS data and NNLO QCD predictions for the transverse sphericity $S_{\perp}$ and aplanarity $A$, both shown in figure \ref{fig:eventshapes-pQCD-2}. Similar level of NNLO/data agreement can also be observed for the $D$ variable shown in figure \ref{fig:eventshapes-pQCD-3}. In terms of the behavior of perturbative corrections, the three variables behave similarly to the thrust. The second-order QCD corrections are most significant in the regions described by events with well-separated and isotropically distributed jets and eliminate the notable shape differences observed at NLO QCD.

The $C$ variable, shown in the left-hand side of figure \ref{fig:eventshapes-pQCD-3}, has a different behaviour. The NLO QCD predictions have little scale dependence in the intermediate region for $C$ and larger scale dependence at the two endpoints $C=0$ and $C=1$. In the central region, neither the normalisation nor the shape of the data is well described within uncertainties. The numerical integration of the NNLO QCD corrections suffers from particularly large cancellations leading to significant remaining statistical errors. For that reason, it is unclear if the second-order corrections improve the theory-data agreement or if the seemingly good agreement with data is simply due to the large statistical errors.

\begin{figure}
    \centering
    \includegraphics[width=\textwidth]{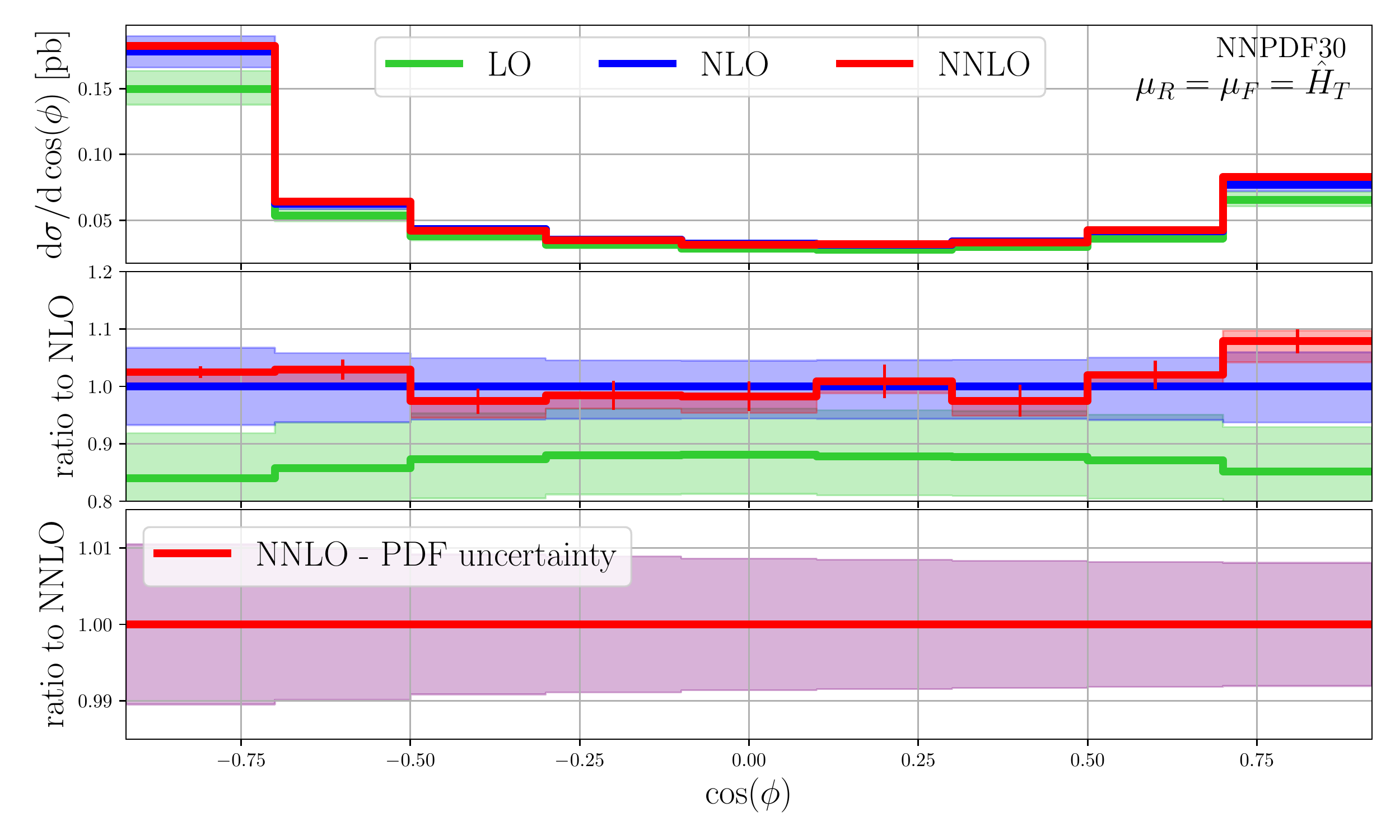}
    \caption{
    The TEEC variable in the inclusive $H_{T,2} \geq 1~\text{TeV}$ bin. The top panel shows the absolute differential distribution through LO (light green), NLO (blue) and NNLO (red) QCD. The coloured bands show the scale uncertainty estimates and vertical bars indicate statistical uncertainties. The second panel shows the ratio to the central NLO QCD prediction. The third panel shows the PDF uncertainty estimate from NNPDF30 at NLO QCD.}
    \label{fig:teec_incl}
\end{figure}

Finally, we turn to the case of the TEEC. In the two top panels in figure \ref{fig:teec_incl}, we show perturbative predictions for the inclusive $H_{T,2}$ selection. A preliminary measurement of this observable has been presented in ref.~\cite{ATLAS:2020mee}. However, the numbers are not yet public and therefore we show no theory/data comparison for this observable in this work.

The typical pattern of higher-order corrections familiar from our discussion of even shapes also emerges for the TEEC. The NLO QCD corrections are sizable relative to the leading order result, barely touching the uncertainty estimate from scale uncertainties. The NLO QCD estimate of the scale uncertainty is flat and about $8\%$. The corrections from NNLO QCD are smaller, between $0\%$ and $5\%$, and lie fully within the NLO scale uncertainty band. The scale variation uncertainty of the NNLO QCD prediction is about $1-3\%$. The remaining numerical integration uncertainty of the NNLO QCD prediction is about $\sim 1-3\%$, i.e. it is similar to the scale uncertainty
\footnote{Since integration uncertainty might lead to imperfect cancellations, the actual scale uncertainty might be smaller than indicated.}. 
Unlike the other event shapes discussed so far, the corrections are relatively flat, and there is no phase space region which receives larger corrections than the others. Notably, within scale uncertainty, the NNLO QCD prediction is consistent with the central NLO QCD one.

\begin{figure}
    \centering
    \includegraphics[page=5,width=\textwidth]{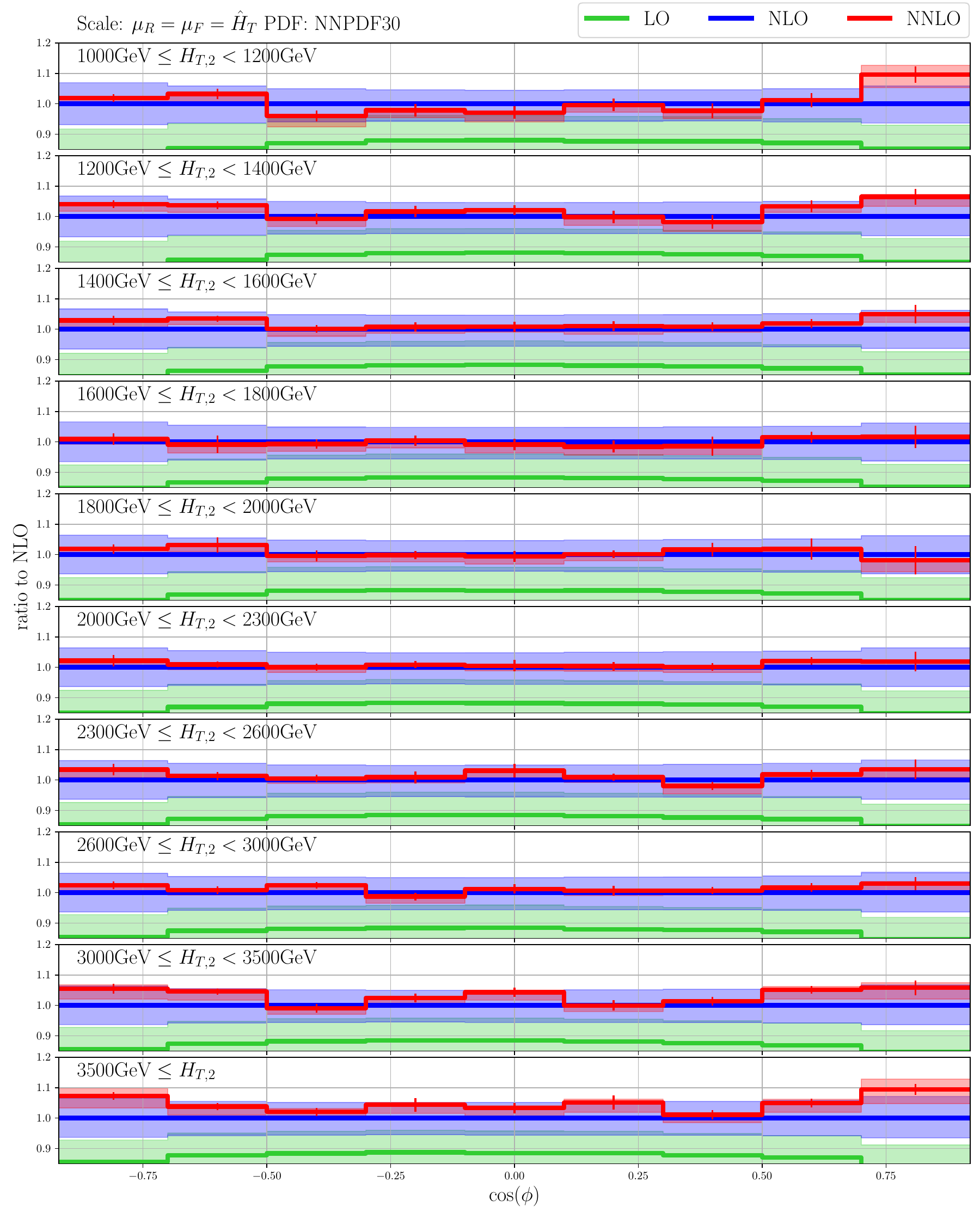}
    \caption{
    The TEEC variable double differentially in $H_{T,2}$ bins. The panels show LO (light green), NLO (blue) and NNLO (red) QCD prediction as a ratio to the central NLO QCD prediction. The coloured bands show the scale uncertainty estimates and vertical bars indicate statistical uncertainties.}
    \label{fig:teec}
\end{figure}

In the double differential case, shown in figure \ref{fig:teec}, we see that the above picture persists also for the individual $H_{T,2}$ slices. Only for $H_{T,2} > 3~\text{TeV}$ we observe an increase in the perturbative correction and scale uncertainty. Still, NNLO QCD corrections remain within the NLO scale uncertainty bands, indicating perturbative stability even at the highest available energies.

%

%

\begin{figure}
    \centering
    \includegraphics[page=1,width=0.5\textwidth]{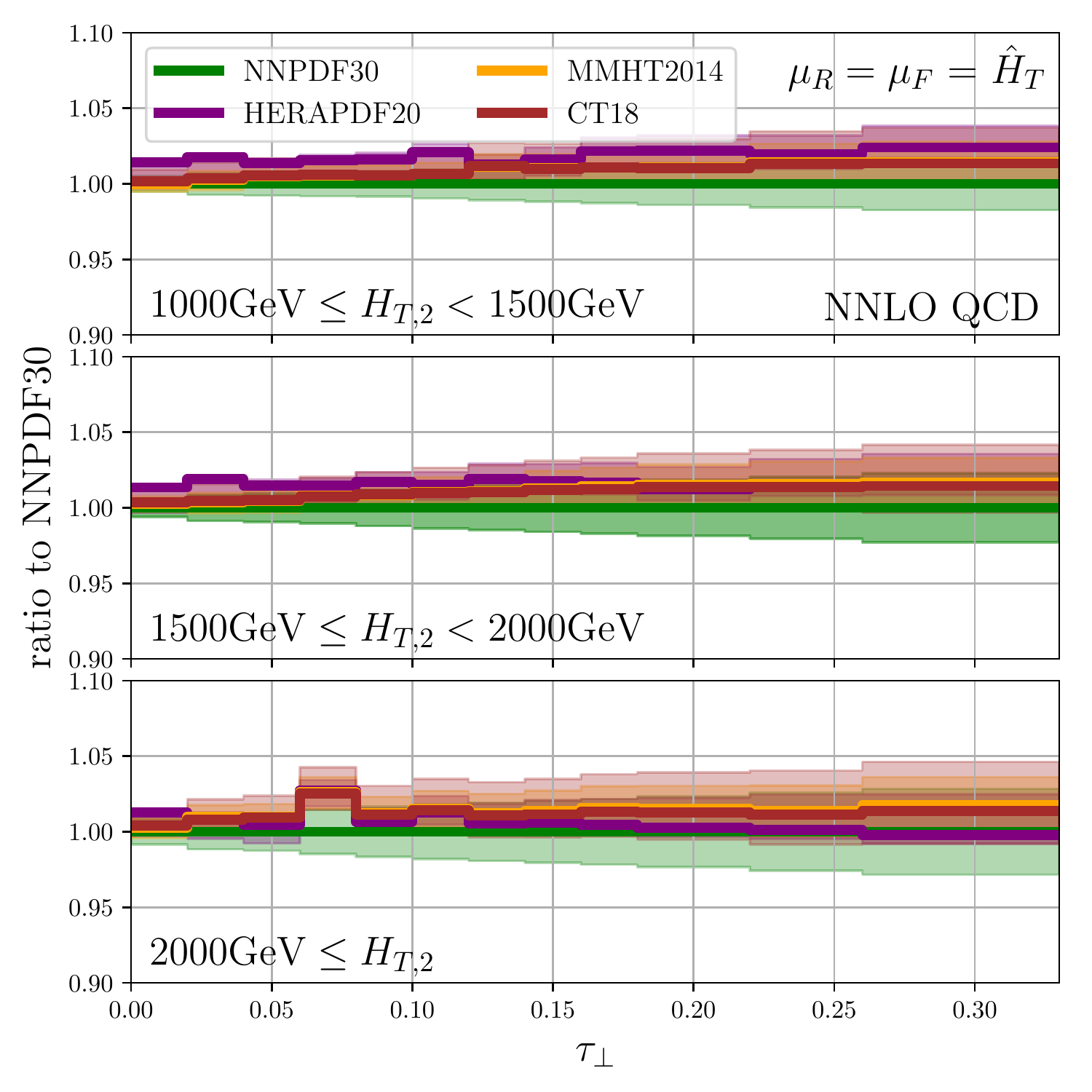}%
    \includegraphics[page=2,width=0.5\textwidth]{plots/NNLO-HThat-r32-pdfs.pdf}
    \caption{
    The transverse thrust $\tau_\perp$ (left) and the thrust minor $T_m$ (right) in three $H_{T,2}$ bins.
    The solid lines show fixed-order NNLO QCD predictions for different PDF sets normalised to NNPDF30.
    The coloured band show the estimated PDF uncertainties.}
    \label{fig:eventshapes-pQCD-pdf}
\end{figure}

We next turn our attention to PDF uncertainties. In fig.~\ref{fig:eventshapes-pQCD-pdf} we show the NNLO QCD predictions for the thrust and the minor component, evaluated with the following PDF sets: \textsc{NNPDF30} \cite{NNPDF:2014otw}, \textsc{MMHT2014} \cite{Harland-Lang:2014zoa}, \textsc{CT18} \cite{Hou:2019qau} and \textsc{HERAPDF} \cite{H1:2015ubc}. The PDF uncertainty estimate for each set is shown as a colored band around the corresponding central value, normalised to the central \textsc{NNPDF30} prediction. The numerical MC uncertainties are not shown because they are fully correlated between different sets, as the same events have been used to evaluate them. 

Starting with the uncertainties for our default PDF set \textsc{NNPDF30}, for both variables we observe that the uncertainties around the peak of the distribution are very small, below $1\%$, and are therefore negligible. The uncertainties increase continuously towards the tails of the distributions and become as large as $2-3\%$. Such a behaviour is expected since the cancellation between PDF dependencies in the numerator and denominator is realised best around the peak of the distribution. The spread between the different PDF sets is about $1-2\%$. All PDFs are compatible within their uncertainty estimates. Furthermore, at NLO QCD the PDF uncertainties are much smaller than the scale ones. At NNLO QCD the scale uncertainties are much reduced -- reaching about $5\%$ in the tails of the distributions -- which leads to PDF uncertainties becoming comparable to the scale ones. Still, at NNLO the scale uncertainty remains the dominant one.

The PDF uncertainties of the other event shapes considered in the present work feature similar behavior. More details can be read off the results that  are available in electronic format \cite{Mitov:URL}.

The PDF uncertainty of the TEEC distribution computed with the \textsc{NNPDF30} PDF set is shown in fig.~\ref{fig:teec_incl} for the inclusive $H_{T,2} > 1~\text{TeV}$ bin. The uncertainty is flat, below $1\%$, indicating that there is a strong cancellation between numerator and denominator for this variable. A similar PDF behavior is also observed in the individual $H_{T,2}$ bins, which we do not show here but supply in electronic form \cite{Mitov:URL}.

\subsection{Strong coupling dependence}\label{sec:eventshapes-alphas}

In this section we quantify the sensitivity of the various event shapes to the value of the strong coupling constant $\aso$. While in the present work we do not attempt to extract a value of $\aso$, the results in this section are an essential step in this direction. To estimate the dependence of the event shapes on the strong coupling constant $\aso$, we evaluate them through NNLO QCD using PDF sets with different values of $\aso$. Each PDF set provides a different range of $\aso$ values, and these have been summarized in table~\ref{tab:alphas_values}. 
\begin{table}
    \centering
    \begin{tabular}{c|c}
         PDF set & $\aso$ values \\
         \hline
         \textsc{NNPDF30}   & $0.115$, $0.117$, $0.118$, $0.119$, $0.121$\\
         \textsc{CT18}      & $0.110$, $0.111$, $\dots$, $0.123$, $0.124$\\
         \textsc{MMHT2014}  & $0.108$, $0.109$, $\dots$, $0.127$, $0.128$\\
         \textsc{HERAPDF20} & $0.110$, $0.111$, $\dots$, $0.129$, $0.130$
    \end{tabular}
    \caption{The values for $\aso = \as(\mu_R=m_Z)$ available with each PDF set.}
    \label{tab:alphas_values}
\end{table}

The $\aso$ dependence in each bin can be derived by expressing the event shape in that bin as a function of $\aso$. This can be achieved as follows. First, consider a perturbative solution of the RGE for the strong coupling constant, i.e.
\begin{align}
    \as (\mu_R,\aso) = \aso\left(1-\aso b_0\ln(\frac{\mu_R^2}{m_Z^2})+\order{\aso^2}\right)\,.
    \label{eq:as-RGE}
\end{align}

Setting for simplicity $\mu = \mu_R = \mu_F$, one can then rewrite eq.~(\ref{eq:ratio_def}) in such a way that the RGE running of the strong coupling is absorbed into the partonic cross section
\begin{align}
    R^{\text{NNLO}}(\mu,\aso) &= \frac{\dd \sigma^{\text{NNLO}}_3(\mu,\aso)}{\dd \sigma^{\text{NNLO}}_2(\mu,\aso)}\nonumber\\
                               &= \frac{\aso^3\left(\dd \tilde{\sigma}^{(0)}_3(\mu) + \aso\dd \tilde{\sigma}^{(1)}_3(\mu) +
                                                   \aso^2\dd \tilde{\sigma}^{(2)}_3(\mu) + {\cal O}(\aso^3)\right)}
                                        {\aso^2\left(\dd \tilde{\sigma}^{(0)}_2(\mu) + \aso\dd \tilde{\sigma}^{(1)}_2(\mu) +
                                                   \aso^2\dd \tilde{\sigma}^{(2)}_2(\mu)+ {\cal O}(\aso^3)\right)}\;.
    \label{eq:ratio_as}
\end{align}

Eq.~(\ref{eq:ratio_as}) makes explicit the non-PDF $\aso$ dependence of the event shape $R^{\text{NNLO}}(\mu,\aso)$. It also makes it clear that the leading $\aso$ dependence of $R^{\text{NNLO}}(\mu,\aso)$ is linear.

Based on the above observations, a very practical way of parameterizing the $\aso$ dependence of the computed event shape is to assume the functional form
\begin{align}
    R^{\text{NNLO,fit}}(\mu,\aso) = c_0 + c_1(\aso-0.118) + c_2(\aso-0.118)^2 + c_3(\aso-0.118)^3\;,
\end{align}
whose coefficients $c_i$ in each bin are determined by fitting $R^{\text{NNLO}}$ as defined in eq.~(\ref{eq:ratio_def})
\footnote{The coefficient $c_0$ is not independent from $c_{1,2,3}$ as follows from the linear $\aso$ dependence of $R^{\text{NNLO}}$.}.
We stress that the fit encodes the unexpanded in $\aso$ value of $R^{\text{NNLO}}$ as well as the $\as$ running obtained directly from the numeric RGE solution as provided by {\tt LHAPDF}, i.e. one does not need to utilize the analytic RGE eq.~(\ref{eq:as-RGE}). 

The coefficients $c_2$ and $c_3$ are typically small. Moreover, one is only interested in percent-level variations of $\aso$ around $0.118$ (which is a convenient proxy for $\aso$'s world average). Therefore, in order to assess the sensitivity of $R^{\text{NNLO}}$ to the value of $\aso$ it is sufficient to focus on the coefficient $c_1$. In practice we consider the rescaled coefficient
\begin{align}
    \tilde{c}_1 = \frac{c_1}{R^{\text{NNLO}}(\aso = 0.118)}\,,
    \label{eq:c1}
\end{align}
which corresponds to the normalised first derivative of $R^{\text{NNLO}}$ with respect to $\aso$ at $\aso = 0.118$. The interpretation of  $\tilde{c}_1$ is that if $\aso$ changes by an amount $\delta \aso$ around $\aso = 0.118$, the ratio $R^{\text{NNLO}}(\aso)/R^{\text{NNLO}}(\aso=0.118)$ changes by $\tilde{c}_1\delta \as$. 

\begin{figure}
    \centering
    \includegraphics[page=1,width=0.5\textwidth]{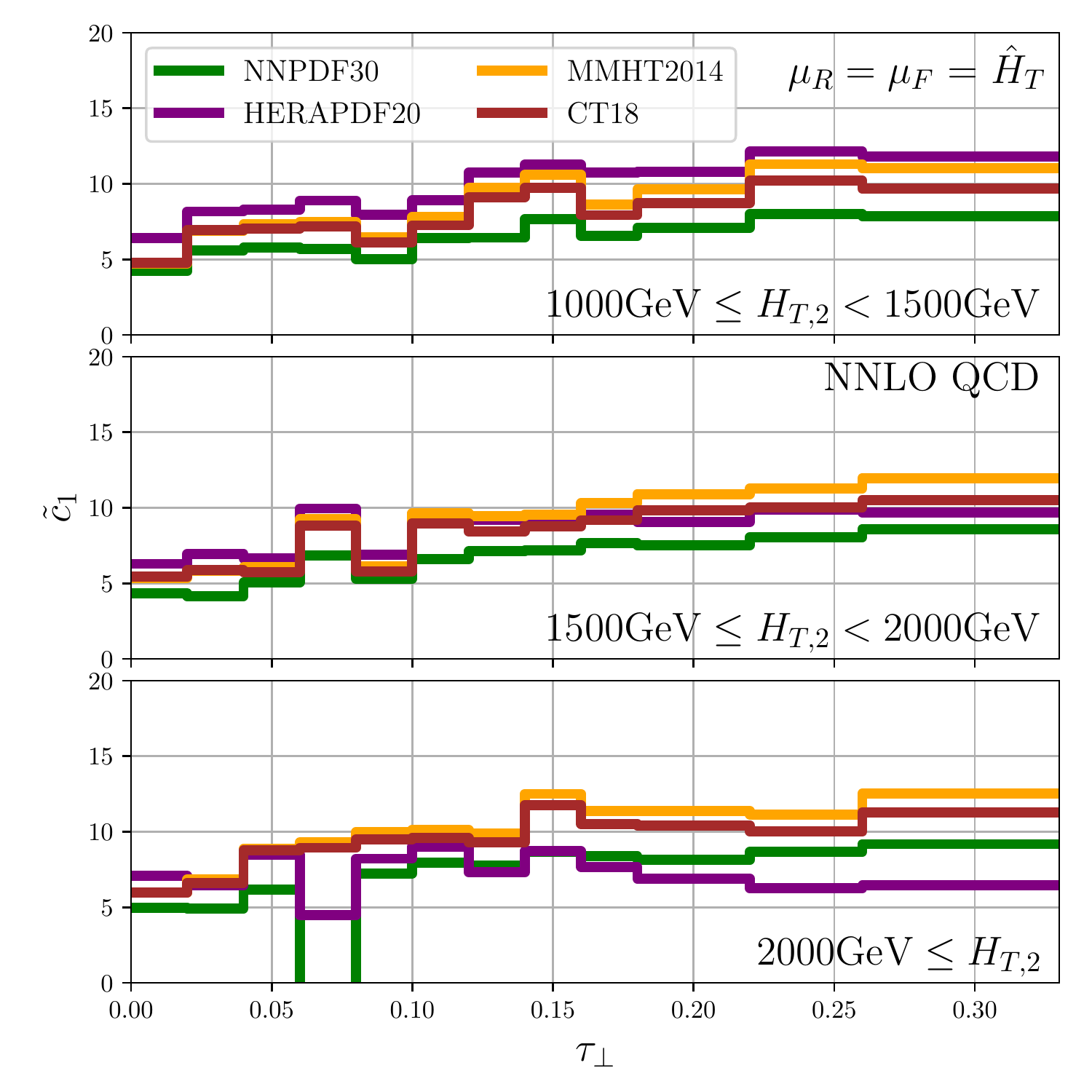}%
    \includegraphics[page=2,width=0.5\textwidth]{plots/NNLO-HThat-r32-alphasdep-rebinned.pdf}
    \caption{
    The transverse thrust $\tau_\perp$ (left) and the thrust minor $T_m$ (right) in three $H_{T,2}$ bins.
    The solid lines show the coefficient $\tilde{c}_1$ at NNLO QCD for different PDF sets.}
    \label{fig:eventshapes-c1}
\end{figure}
\begin{figure}
    \centering
    \includegraphics[page=2,width=0.8\textwidth]{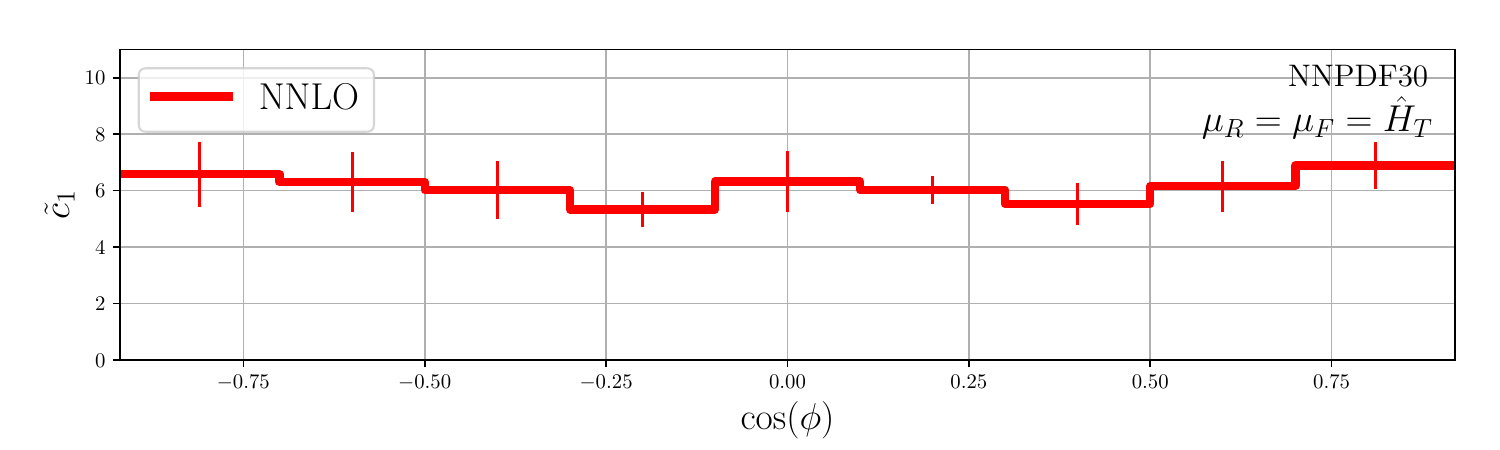}
    \caption{As in fig.~\ref{fig:eventshapes-c1} but for the TEEC.}
\label{fig:TEEC-c1}
\end{figure}

The NNLO value of the coefficient $\tilde{c}_1$ for the event shapes $\tau_\perp$ and $T_m$ is shown in fig.~\ref{fig:eventshapes-c1}. We observe that the relative sensitivity of these two observables to $\aso$ increases with larger values of the corresponding kinematic variables, reaching a plateau of $\tilde{c}_1 \approx 10$. This increase in sensitivity is consistent with the observation that the contribution from higher multiplicity matrix elements becomes more significant for larger thrust values. These come with additional powers of $\as$, increasing the sensitivity to $\aso$. A value $\tilde{c}_1 \approx 10$ implies that a $1\%$ shift in $\aso$ leads to $\approx 1\%$ shift in the prediction. This suggests that these event shapes are suitable for potential future extraction of $\aso$ from LHC data.
 
In fig.~\ref{fig:TEEC-c1} we show the dependence of the TEEC on $\aso$. The sensitivity of this observable is largely kinematics-independent, with a value $\tilde{c}_1 \approx 6$. This implies that a $1\%$ shift in $\aso$ changes the TEEC by about $0.6\%$, which is similar to the $\aso$ sensitivity of the other event shapes around their peak regions.

Up to this point we considered the case where one uses the measured event shapes in order to extract $\aso$, i.e. the value of $\as$ at the standard reference scale $\mu_F=m_Z$. Besides this reference value of the strong coupling, there is a long-standing interest in experimentally verifying its SM running, i.e. the dependence of $\as$ on its energy scale. Keeping in mind that $\as$ is {\it not an observable}, which implies that there is an arbitrariness in the energy scale that is associated with a given measurement. Technically, $\as$ is ran up to the scale $\mu_R$, therefore, a natural choice for the $\as$ scale associated to a binned measurement is the mean value $\langle\mu_R\rangle$ of the renormalisation scale in each bin. In this work we will not elaborate on the question of choosing this scale (see, for example, ref.~\cite{Czakon:2016dgf} for a broader discussion on this topic). We will only remark that if a particular choice of $\mu_R$ results in a perturbatively convergent prediction, it is reasonable to assume that it represents the relevant physical scale for this process and observable. In this work we use the scale choice $\mu_R = \hat{H}_T$ which, indeed, exhibits good perturbative convergence. From this we conclude that the best choice for the energy scale in each $H_{T,2}$ bin is $\langle \hat{H}_T\rangle$ and in the context of our calculation, it should be interpreted as the energy scale at which $\as$ is extracted. The results for $\langle \hat{H}_T\rangle$ in each $H_{T,2}$ bin is shown in table~\ref{tab:eventshapes-avg-hth}.

\begin{table}[t]
    \centering
    \begin{tabular}{c|c|c|c}
         $H_{T,2}$ bin               & $[1000,1500]$ GeV & $[1500,2000]$ GeV & $\geq 2000$ GeV  \\
         \hline
         $\langle\hat{H}_{T}\rangle$ & $1371 \pm 7$ GeV  & $1928 \pm 13$ GeV & $2607 \pm 15$ GeV
    \end{tabular}
    \caption{The average $\hat{H}_{T}$ in each $H_{T,2}$ bin computed from the integrated cross section of $\dd\sigma_2^{\text{NNLO}}$ in each bin. The uncertainty indicated above is from MC integration.}
    \label{tab:eventshapes-avg-hth}
\end{table}

\subsection{Estimation of non-perturbative corrections}\label{sec:eventshapes-np}

\begin{figure}
    \centering
    \includegraphics[page=1,width=0.5\textwidth]{{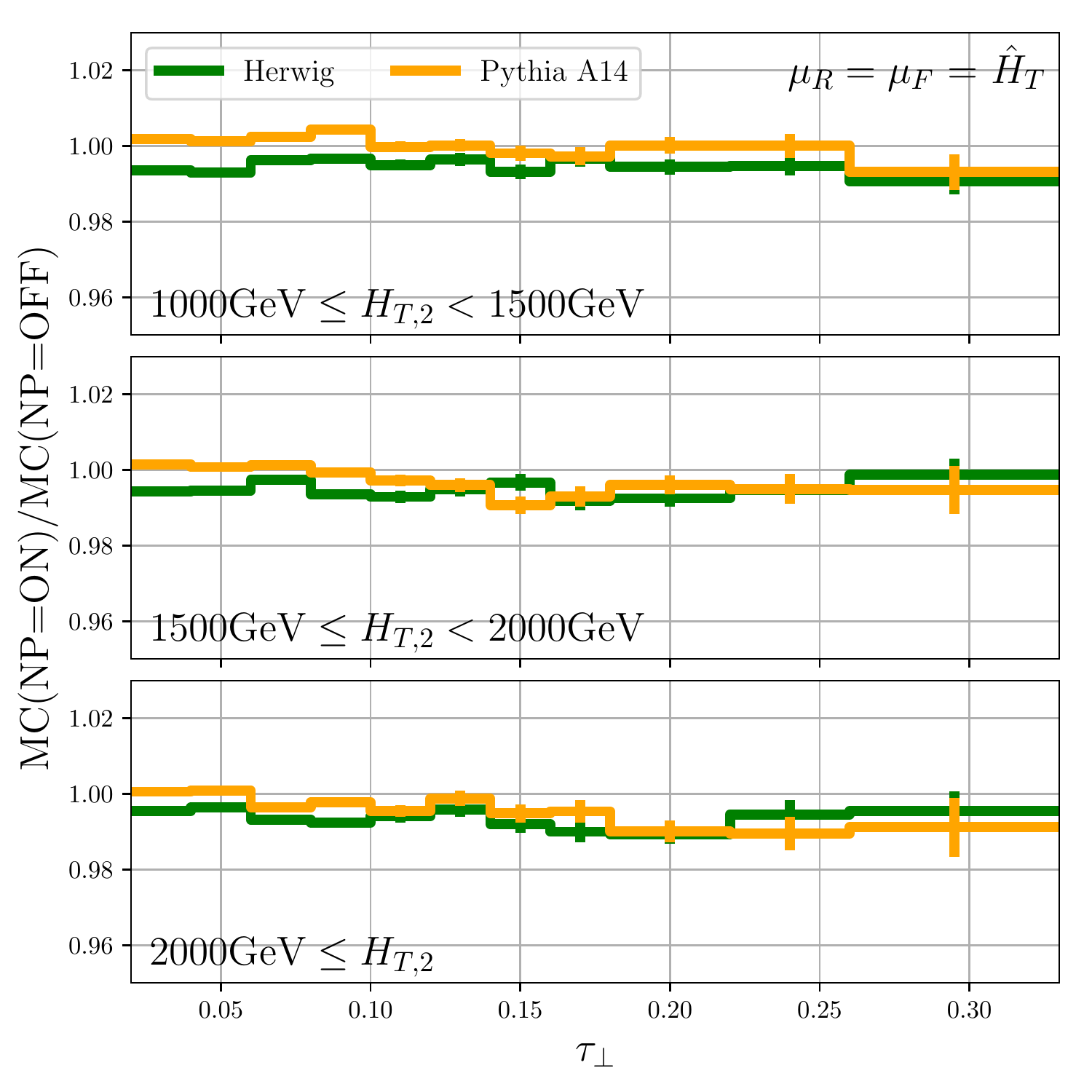}}%
    \includegraphics[page=2,width=0.5\textwidth]{{plots/np-r32-rebinned.pdf}}
    \caption{
    The transverse thrust $\tau_{\perp}$ (left) and the thrust minor $T_m$ (right) in three $H_{T,2}$ bins.
    The solid lines show the estimated non-perturbative effects from \textsc{Herwig} and \textsc{Pythia}.}
    \label{fig:eventshapes-np}
\end{figure}

We conclude the discussion of event shape observables with a discussion of non-perturbative effects. Non-perturbative corrections can have a non-trivial impact on event shapes, see for example the discussion of thrust \cite{Becher:2008cf} and energy-energy correlator \cite{Tulipant:2017ybb} in $e^+e^-$ collisions, and can impact the accuracy of $\as$ extractions. The observables studied in this work are based on clustered jets which reduces the sensitivity to such effects. One possibility for assessing the effects of hadronisation and multi-parton interactions (MPI) is to make use of Monte Carlo event generators. For this purpose, we evaluate the event shapes with \textsc{Herwig} \cite{Bahr:2008pv, Bellm:2015jjp, Bellm:2017bvx} and \textsc{Pythia} \cite{Sjostrand:2006za, Sjostrand:2014zea} at LO
\footnote{NLO+PS is not yet readily available for three jet production. The exception is a study with the \textsc{Sherpa} event generator~\cite{Reyer:2019obz}.}, 
once with active hadronisation and MPI and once without. The ratio of the two predictions serves as an estimate of the expected non-perturbative corrections. Numerical results for the thrust and thrust minor are presented in figure \ref{fig:eventshapes-np}. The non-perturbative effects reach $1\%$ which is rather small. This confirms the expectation that for the event shapes considered in this work the non-perturbative effects are subdominant to other sources of theory uncertainty.

\section{Conclusions}\label{sec:conclusion}

In this work we perform the first calculation of jet event shapes at hadron colliders at NNLO in QCD. Specifically, we consider the transverse thrust $\tau_{\perp}$ and its minor component $T_m$, the shapes $A$, $C$, and $D$ derived from the linearised sphericity tensor, the transverse sphericity variable $S_\perp$ and the transverse energy-energy correlator. In order to be able to describe the full kinematics of these event shapes, one needs to include all contributions with at least three jets in the final state. Such a calculation became possible at NNLO only very recently \cite{Czakon:2021mjy}. 

The immediate goal of this work is to clarify if higher order corrections to jet event shapes at hadron colliders significantly improve the theory/data comparison for these observables. Another goal relates to the well-known fact that such observables require all-order resummation in certain regions of phase space. One would like to clarify if by including NNLO QCD corrections to these observables, kinematic regions where fixed order perturbation theory is reliable can be clearly identified. Such regions are important because in $e^+e^-$ collisions they are typically used for measuring the strong coupling constant as well as for tuning parameters of shower Monte Carlo event generators. 

In this work we provide predictions for typical ATLAS setups at 13 TeV and compare them with data where available (only for the TEEC no public numbers are available). Across all event shapes we observe  that NNLO QCD reduces significantly the scale uncertainty of the predictions, typically by a factor of 2 to 4 relative to NLO. More importantly, the inclusion of NNLO QCD corrections has a large impact on the shapes of these observables. At NLO QCD one typically observes a theory/data agreement within the theory scale uncertainty, however, the shapes of theory and data tend to be rather different. Once NNLO corrections are included the theory and data shapes tend to ``align" well. 

Once NNLO QCD corrections are included, one can clearly identify narrow kinematic regions where fixed order predictions are unreliable. Likely, this is due to missing all-order resummation. As it might be expected, for all observables that show such a behavior, this is the limit where a three-jet final state starts to resemble a two-jet one. Outside of this relatively narrow region, the event shapes are reliably described by fixed order calculations. 

For all event shapes we observe that the total experimental uncertainty tends to be smaller than the theory one at NNLO QCD. The dominant source of theory uncertainty is scale variation. A second, comparable source of theory uncertainty is the Monte Carlo integration one. The three jet calculation is extremely computationally expensive and one cannot expect to improve on it unless very significant computational resources are deployed. This might be required for future high precision theory/data comparisons, since in places where the MC uncertainty is large it tends to also blow up the estimated scale variation. A smaller but not insignificant source of theoretical uncertainty is the PDF one. We estimate it to be probably about half the scale one or less, which makes it less relevant in immediate precision applications. We have also estimated the effect of non-perturbative corrections which we find to be around or below 1\%, and therefore negligible. We have not investigated the effect of EW corrections. These are expected to be small, partly because the event shapes are defined as ratios of three- to two-jet cross sections.

Important immediate applications of our results relate to the extraction of the strong coupling constant. An obvious application is the precision determination of $\as(m_Z)$ from LHC jet data. Although we do not perform such an extraction in this work, we have provided a detailed investigation of its feasibility and prospects. Our analysis demonstrates that the event shapes considered in the present work have sensitivities to the value of $\as(m_Z)$ of between about $0.5\%$ and $1\%$.  This makes them suitable for the extraction of $\as$ in the kinematic regions where fixed order perturbation theory is reliable. A second, no less important, application is the measurement of the running of $\as$. The suitability of the three-to-two jet cross section for such a measurement has been well known for a very long time, however the readily available NLO QCD predictions \cite{Nagy:2001fj, Nagy:2003tz} are not precise enough for performing such a measurement. The calculation of the NNLO QCD corrections, provided in the present work, allows for the first time to precisely map out the running of the QCD coupling constant to energy scales as large as several TeV. A measurement with such an unprecedented precision will allow new ways for searching for physics beyond the SM and for improving our understanding of the running of the SM coupling constants well above the EW scale.

\begin{acknowledgments}
We would like to thank Andrea Banfi for insightful discussions about event shapes.
The work of M.C. was supported by the Deutsche Forschungsgemeinschaft under grant 396021762 – TRR 257. The research of A.M. and R.P. has received funding from the European Research Council (ERC) under the European Union's Horizon 2020 Research and Innovation Programme (grant agreement no. 683211). A.M. was also supported by the UK STFC grants ST/L002760/1 and ST/K004883/1. R.P. acknowledges support from the Leverhulme Trust and the Isaac Newton Trust.
This work was performed using the Cambridge Service for Data Driven Discovery (CSD3), part of which is operated by the University of Cambridge Research Computing on behalf of the STFC DiRAC HPC Facility (www.dirac.ac.uk). The DiRAC component of CSD3 was funded by BEIS capital funding via STFC capital grants ST/P002307/1 and ST/R002452/1 and STFC operations grant ST/R00689X/1. DiRAC is part of the National e-Infrastructure. Simulations were performed with computing resources granted by RWTH Aachen University under project p0020025.
\end{acknowledgments}

\bibliography{main}
\bibliographystyle{JHEP}

\end{document}